\documentclass[aps,prb,showpacs,superscriptaddress,twocolumn,final]{revtex4}
\usepackage{graphicx}
\usepackage{epsfig}
\usepackage[american]{babel}
\usepackage{color}
\usepackage{textcomp}
\usepackage{amsmath}
\usepackage{amssymb}
\usepackage{amsfonts}
\usepackage[hang,normalsize,sf,SF]{subfigure}

\newcommand{\un}[1]{\,\mathrm{#1}}

\newcommand{\mvektor}[1]{\ensuremath{\mathbf{#1}}}%
\newcommand{\mmatrix}[1]{\ensuremath{\mathfrak{#1}}}%

\newcommand{\VSD}{\ensuremath{V_\text{SD}}}

\newcommand{\goodgap}{
 \hspace{\subfigtopskip}%
 \hspace{\subfigbottomskip}}

\begin{document}

\title{An electrostatically defined serial triple quantum dot charged with few electrons}

\author{D.\ Schr\"oer}
\affiliation{Center for NanoScience and Department f\"ur Physik,
Ludwig--Maximilians--Universit\"at, Geschwister--Scholl--Platz 1,
80539~M\"unchen, Germany.}

\author{A.\,D.\ Greentree}
\affiliation{Centre for Quantum Computer Technology, School of Physics,
University of Melbourne, Victoria 3010, Australia.}

\author{L.\ Gaudreau}
\affiliation{Center for NanoScience and Department f\"ur Physik,
Ludwig--Maximilians--Universit\"at, Geschwister--Scholl--Platz 1,
80539~M\"unchen, Germany.}
\affiliation{Institute For Microstructural Sciences, NRC, Ottawa, Canada K1\,A\,0R6.}
\affiliation{R\'egroupement Qu\'eb\'ecois sur les Mat\'eriaux de Pointe, Universit\'e de Sherbrooke, Canada J1K 2R1.}

\author{K.\ Eberl}
\altaffiliation[Present address: ]{Lumics GmbH, Carl--Scheele--Strasse
  16, 12489 Berlin, Germany.}
\affiliation{Max-Planck-Institut f\"ur Festk\"orperforschung,
Heisenbergstra{\ss}e 1, 70569 Stuttgart, Germany.}

\author{L.\,C.\,L.\ Hollenberg}
\affiliation{Centre for Quantum Computer Technology, School of Physics,
University of Melbourne, Victoria 3010, Australia.}

\author{J.\,P.\ Kotthaus}
\affiliation{Center for NanoScience and Department f\"ur Physik,
Ludwig--Maximilians--Universit\"at, Geschwister--Scholl--Platz 1,
80539~M\"unchen, Germany.}

\author{S.\ Ludwig}
\email{Stefan.Ludwig@Physik.Uni-Muenchen.de}
\affiliation{Center for NanoScience and Department f\"ur Physik,
Ludwig--Maximilians--Universit\"at, Geschwister--Scholl--Platz 1,
80539~M\"unchen, Germany.}

\date{\today}


\begin{abstract}
A serial triple quantum dot (TQD) electrostatically defined in a GaAs/AlGaAs heterostructure is characterized by using a nearby quantum point contact as charge detector. Ground state stability diagrams demonstrate control in the regime of few electrons charging the TQD. An electrostatic model is developed to determine the ground state charge configurations of the TQD. Numerical calculations are compared with experimental results. In addition, the tunneling conductance through all three quantum dots in series is studied. Quantum cellular automata processes are identified, which are where charge reconfiguration between two dots occurs in response to the addition of an electron in the third dot.
\end{abstract}

\maketitle

\section{Introduction}

Extensive experimental work has recently been aimed towards electrostatically defining and controlling semiconductor quantum dots (QDs) and double quantum dots (DQD).\cite{Kou97,ciorga,Hof95,Wie03,Elz03,prl-hayashi:226804,Pio05,Hue05,science-petta:2180,nature-koppens:766,Hue07} The complete control of the QD charge, down to the limit of only one trapped conduction band electron, has been demonstrated by monitoring the single electron tunneling current through the device\cite{ciorga}, by counting the charge on the QD electron by electron by means of a nearby quantum point contact (QPC)\cite{Elz03,Pio05} or by combining both methods\cite{Hue07}. Such efforts are predominantly motivated by the desire to control and understand the physics of quantum systems, and provide impetus for proposals for using the spin\cite{pra-loss:120}, charge states\cite{jjap-vanderwiel:2100} or encoded subspaces\cite{DiVi00,Tay05} of localized electrons as qubits, the elementary registers of the hypothetical quantum computer. Recent experiments have demonstrated the realization and coherent control of charge\cite{prl-hayashi:226804} and spin qubits\cite{science-petta:2180, nature-koppens:766} in DQDs.

Extending (double) QD circuits towards a few electron triple quantum dot (TQD) is a natural step towards scalable multi-qubit systems. In addition, the spin states in three tunnel-coupled QDs can be used to encode a qubit in the logical states of a decoherence-free subspace. In this way the coherence time of the qubit is expected to increase and gate operations to be simplified at the cost of a higher number of required QDs.\cite{DiVi00,Sas04,Haw05}

Quantum information processing relies on a coupling between qubits allowing coherent exchange of quantum information\cite{pra-loss:120}. In QD-based implementations of quantum computing, where qubit coupling is local, introducing coherent qubit transport is important in the design of a scalable fault-tolerant architecture.\cite{Hol06} Coherent transfer by adiabatic passage (CTAP) has been proposed as a way to efficiently move electrons along chains of tunnel coupled QDs and entangle quantum mechanical states of distant qubits.\cite{Gre04,Gre05} A TQD is the smallest system that in principle allows the implementation of CTAP.

In addition to applications in quantum information processing, the interest in TQDs is triggered by a rich spectrum of phenomena going beyond the physics of DQDs. These include combined charging and reconfiguration events that can be identified as quantum cellular automata (QCA) processes,\cite{Len93} applications as current rectifiers,\cite{Sto02,Vid04} creation of spin-entangled electrons,\cite{Sar03,Zha04} and new aspects of the Kondo\cite{Kuz02, Kuz06, Zit06} and Fano\cite{Lad06} effects.

Several efforts have been undertaken to produce laterally defined TQDs. In an early attempt, large TQDs in a serial configuration were studied via transport measurements as a function of the coupling between the QDs.\cite{Wau95, Wau96} More recently, current rectification effects were observed in devices consisting of three tunnel coupled QDs charged with many electrons.\cite{Sto02, Vid04, Vid05} Charge stability diagrams at low electron numbers were first investigated in a geometry in which one of two coupled QDs is split further, thus realizing a TQD in a ring-like device consisting of three tunnel coupled QDs.\cite{Gau06} The mapping of charge stability diagrams revealed a QCA effect near points of resonant transport. Magneto-conductance experiments further unveiled Aharonov-Bohm like oscillations.\cite{Gau07} The realization of three laterally coupled vertical quantum dots is under investigation.\cite{Kim06}

In this article we report on the realization of a TQD in a serial configuration charged with few electrons. The gate layout was specifically designed to define three small QDs tunnel coupled in series. We characterize the TQD by means of stability diagrams. Integrated charge detection is performed using a nearby QPC. In addition, electron tunneling transport through the three QDs in series is investigated. Observed features, which are specific for a TQD, including triple points, quadruple points, and QCA effects, are discussed in detail. We derive a classical electrostatic model that allows us to predict charge stability diagrams of a TQD by minimizing its free energy. The model is easily scalable to larger systems containing more QDs. A detailed comparison between this model and our data is presented in a regime close to points of sequential resonant transport through the TQD.

\section{The triple quantum dot layout}

Our sample is fabricated from an AlGaAs/GaAs heterostructure with a two-dimensional electron system (2DES) embedded $120\un{nm}$ below the surface.  At $T=4.2\un{K}$ the 2DES features an electron sheet density of $n_\text{s} \simeq 1.8\times 10^{15}\un{m}^{-2}$ and a mobility of $\mu \simeq 75 \un{m}^2/\text{Vs}$. Experiments are performed in a dilution refrigerator at an electron temperature of $T_\text{2DES} \simeq 100\un{mK}$, as determined by the width of Coulomb blockade conductance peaks.\cite{Gol98}

\begin{figure}[th]
\begin{center}
\epsfig{file=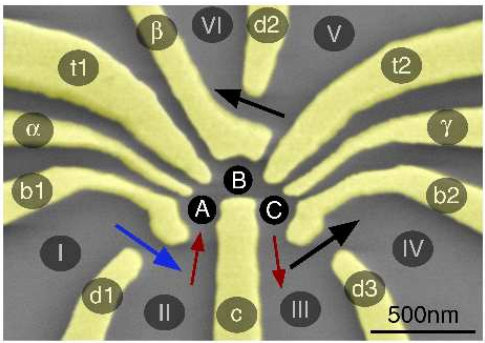, width=6.5cm}
\end{center} 
\caption{\label{fig1}(Color online) SEM micrograph of the sample structure. Gate electrodes (bright tone) are used to electrostatically define a TQD and three QPCs. The approximate position of the three QDs (A, B, and C) are depicted by black circles. Large (small) arrows mark possible tunneling current paths through QPCs (the TQD). Ohmic contacts are labeled with roman numbers. The gates marked with $\alpha$, $\beta$, and $\gamma$ are used as plunger gates of the three QDs A, B, and C. Gates marked as d1, d2, and d3 serve to define QPCs as charge sensors.}
\end{figure}

Electron beam lithography is used to produce Ti/Au gates on the surface of the heterostructure as shown in Fig.~\ref{fig1}.
The TQD and up to three QPCs are defined by applying appropriate negative voltages to the gates to locally deplete the 2DES beneath. The gate layout extends a single QD geometry that allows transport spectroscopy at low electron numbers.\cite{ciorga} Our sample allows the definition of up to three QDs ( A, B, and C as indicated by black circles in Fig.~\ref{fig1}) tunnel coupled in a serial configuration. Transport measurements can be performed even in the regime of only few electrons charging the QDs. Three independent QPCs (marked with arrows in Fig.~\ref{fig1}) can be used to determine the charge configuration of the TQD in the same way as has been demonstrated for the case of DQDs.\cite{Elz03,Pio05,Hue05} The described approach using laterally defined surface gates is in principle scalable to much larger systems containing many QDs.

\section{Electrostatic model of a triple quantum dot stability diagram}

A charge stability diagram affords a quick and intuitive mechanism to understand many of the properties of a quantum electronic system. As the surface gate voltages are varied, the system tries to minimize its free energy by exchanging electrons with the leads and by redistributing the charges between its constituents. In the case of a DQD the stable charge configurations form a characteristic honeycomb diagram as a function of the voltages applied to two plunger gates.\cite{Elz03,Pio05} The TQD introduces further complexity and richness of phenomena. The obvious choice for a full description of all possible charge configurations of a TQD would be a three-dimensional stability diagram as a function of three plunger gate voltages. Here, we investigate two-dimensional slices of such a three-dimensional stability diagram.

\begin{figure}[th]
\begin{center}
\epsfig{file=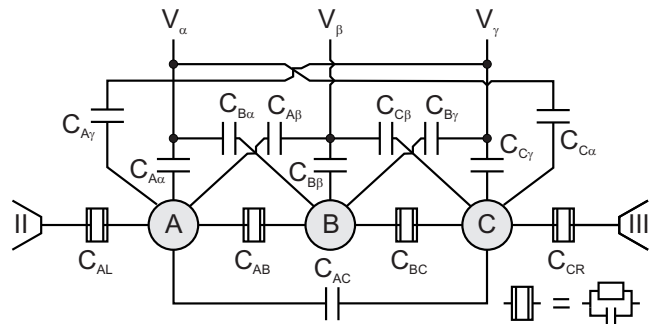, width=8.5cm}
\end{center}
\caption{\label{fig2} Equivalent circuit diagram for three tunnel coupled QDs A, B, and C in serial configuration. QDs A and C are, in addition, tunnel coupled to leads II and III, respectively. Tunnel barriers are modeled as resistors and capacitors in parallel and electrostatic coupling to three plunger gates $\alpha$, $\beta$, and $\gamma$ as capacitors.}
\end{figure}

Standard electrostatic models describing a DQD\cite{Hof95,Wie03} can be extended towards a TQD.\cite{Was97,Vid05} We introduce a scalable matrix approach describing electrostatic Coulomb-interaction by capacitance matrices. Quantum mechanical tunneling between QDs is not taken explicitly into account for the classical model. Implicitly, tunneling of electrons allows transitions between charge configurations. Fig.~\ref{fig2}
sketches an equivalent circuit diagram for a serial TQD. It models tunnel barriers of the TQD as ohmic resistors and capacitors in parallel and electrostatic coupling to three plunger gates as capacitors. During a typical measurement all other gate voltages are kept constant. They are not included in the equivalent circuit diagram in Fig.~\ref{fig2} for simplicity.
The relevant circuit consists of charge nodes (QDs A, B, and C), voltage nodes (plunger gates $V_\alpha$, $V_\beta$, and $V_\gamma$), and capacitors separating nodes. The electrostatic potential of the 2DES including source and drain leads is assumed to be at ground level (i.\,e.\ $V_{II}=V_{III}=0$). This is a good approximation for a typical measurement in the linear response regime.

In a structure with $N$ nodes with electrostatic potentials $V_j$, we can express the total charge $Q_j$ of each node (including voltage nodes) as
\begin{eqnarray*}
Q_j = \sum_{k=1}^N q_{jk} = \sum_{k=1}^N C_{jk}\left(V_j-V_k\right),
\end{eqnarray*}
where $q_{jk}$ and $C_{jk}$ are the continuous polarization charge and capacitance between nodes $j$ and $k$. This expression is more conveniently written in matrix notation as  $\mvektor{Q} = \mmatrix{C} \mvektor{V}$, where $\mvektor{Q}$ and $\mvektor{V}$ are vectors with the elements $Q_j$ and $V_j$, respectively. The matrix $\mmatrix{C}$ contains the capacitances $C_{jk}$ between nodes.
The diagonal matrix elements $C_{jj}\equiv C_j^{\Sigma}$ are the self capacitances of nodes $j$, defined as the sum of the capacitances between the node and all other objects in the universe. Explicitly, our model only considers variable voltages applied to the plunger gates. All other gates have constant electric potentials but contribute implicitly via the self capacitances $C_j^{\Sigma}$.

We can separate the matrix equation $\mvektor{Q} = \mmatrix{C} \mvektor{V}$ as
\begin{eqnarray}
\left[%
\begin{array}{c}
  \mvektor{Q_\text{D}}\\
  \mvektor{Q_\text{V}}\\
\end{array}%
\right] = \left[%
\begin{array}{cc}
  \mmatrix{C_\text{DD}} & \mmatrix{C_\text{DV}} \\
  \mmatrix{C_\text{DV}}^{\text{T}} & \mmatrix{C_\text{VV}} \\
\end{array}%
\right]\left[%
\begin{array}{c}
  \mvektor{V_\text{D}}\\
  \mvektor{V_\text{V}}\\
\end{array}%
\right],
\label{eq2}
\end{eqnarray}
where charge nodes (QDs) are labeled with subscript D, and voltage nodes (gates) with V. The vectors $\mvektor{Q_\text{D}} = \left[Q_\text{A},Q_\text{B},Q_\text{C}\right]^\text T$, $\mvektor{Q_\text{V}} = \left[Q_\alpha,Q_\beta,Q_\gamma\right]^\text T$, $\mvektor{V_\text{D}} = \left[V_\text{A},V_\text{B},V_\text{C}\right]^\text T$, and $\mvektor{V_\text{V}} = \left[V_\alpha,V_\beta,V_\gamma\right]^\text T$ contain the total charges and voltages on the three QDs and three gates, respectively. The matrix $C$ is split into
\begin{displaymath}
\mmatrix{C_\text{DD}} = \left[%
\begin{array}{ccc}
  C_\text{A}^{\Sigma} & -C_\text{AB} & -C_\text{AC}\\
  -C_\text{AB} & C_\text{B}^{\Sigma} & -C_\text{BC}\\
  -C_\text{AC} & -C_\text{BC} & C_\text{C}^{\Sigma}
\end{array}%
\right],
\end{displaymath}
containing only capacitances between QDs and the self capacitances of the QDs,
\begin{displaymath}
\mmatrix{C_\text{DV}} = \left[%
\begin{array}{ccccc}
  -C_{\text A\alpha} & -C_{\text A\beta} & -C_{\text A\gamma}\\
  -C_{\text B\alpha} & -C_{\text B\beta} & -C_{\text B\gamma}\\
  -C_{\text C\alpha} & -C_{\text C\beta} & -C_{\text C\gamma}
\end{array}%
\right],
\end{displaymath}
and its transpose containing all capacitances between gates and QDs, and \mmatrix{C_\text{VV}} containing all capacitances between the three plunger gates.

In our experiments the electrostatic potentials on all gates are independent of the capacitances between the gates, because they are imposed by externally applied voltages. Hence, the matrix \mmatrix{C_\text{VV}} only influences the zero point of energy. For simplicity we assume $\mmatrix{C_\text{VV}}=0$ without loss of generality.

Our intention is, to find the ground state stability diagram of a TQD by numerically minimizing its free energy $F=U-W$. Here, $U$ is the electrostatic energy of a given configuration created by achieving the work $W$. It is useful to introduce the total effective charge of the QDs defined as the sum of $\mvektor{Q}_\text D$ and the electrostatic influence charge $- \mmatrix{C_\text{DV}} \mvektor{V}_\text{V}$. The relevant part of eq.~\ref{eq2} then reads
\begin{eqnarray}
\mvektor{Q}_{\mathrm{D}}^{\mathrm{eff}} \equiv \mmatrix{C_\text {DD}} \mvektor{V}_\text {D} &=& \mvektor{Q}_{\mathrm{D}} - \mmatrix{C_\text {DV}} \mvektor{V}_\text {V}\nonumber\\ &=& e\mvektor{N}_{\mathrm{D}} - \mmatrix{C_\text {DV}} \mvektor{V}_\text {V}\nonumber,
\end{eqnarray}
where we allow only discrete values of the charges of the QDs expressed by $\mvektor{Q}_\text D = e \mvektor{N}_\text D$ with the electronic charge $e$. The vector $\mvektor{N}_\text D = [N_\text A, N_\text B, N_\text C]^{\mathrm{T}}$ contains the number of electrons per QD and defines the charge configuration ($N_\text A,N_\text B,N_\text C$) of the TQD. The free energy reads
\begin{eqnarray}
F &=& U-W\nonumber\\
&=& \frac 1 2
\left[\mvektor{Q_\text{D}}^{\text{T}},\mvektor{Q_\text{V}}^{\text{T}}\right]
    \left[
        \begin{array}{c}
        \mvektor{V}_{\mathrm{D}} \\
        \mvektor{V}_{\mathrm{V}} \\
        \end{array}
    \right]
 - \mvektor{V_\text{V}}^{\text{T}}\mvektor{Q_\text{V}}\nonumber\\
&=& \frac 1 2 \left( \mmatrix{C_\text{DD}}^{-1} \mvektor{Q_\text{D}}^\text{eff} \right)^\text{T} \mvektor{Q_\text{D}}^\text{eff}\nonumber\\
& = & \frac 1 {2e^2} Q_\text{A}^\text{eff} \left( E_\text{A} Q_\text{A}^\text{eff} + E_\text{AB} Q_\text{B}^\text{eff} + E_\text{AC} Q_\text{C}^\text{eff} \right)\nonumber\\
&+& \frac 1 {2e^2} Q_\text{B}^\text{eff} \left( E_\text{AB} Q_\text{A}^\text{eff} + E_\text{B} Q_\text{B}^\text{eff} + E_\text{BC} Q_\text{C}^\text{eff} \right)\nonumber\\
&+& \frac 1 {2e^2} Q_\text{C}^\text{eff} \left( E_\text{AC} Q_\text{A}^\text{eff} + E_\text{BC} Q_\text{B}^\text{eff} + E_\text{C} Q_\text{C}^\text{eff} \right),\label{free_energy}
\end{eqnarray}
where
\begin{eqnarray*}
Q_X^\text{eff} &=& Q_X + C_{X\alpha} V_\alpha + C_{X\beta} V_\beta + C_{X\gamma} V_\gamma\\
E_X &=& K\left( C_{Y}^{\Sigma} C_{Z}^{\Sigma} - C_{YZ}^2 \right)\\
E_{XY} &=& K\left( C_{Z}^{\Sigma} C_{XY} + C_{XZ} C_{YZ} \right)\\
K &=& e^2/ (C_\text{A}^{\Sigma} C_\text{B}^{\Sigma} C_\text{C}^{\Sigma} - 2 C_\text{AB} C_\text{AC} C_\text{BC}\\
  &&- C_\text{C}^{\Sigma} C_\text{AB}^2 - C_\text{B}^{\Sigma} C_\text{AC}^2 - C_\text{A}^{\Sigma} C_\text{BC}^2),
\end{eqnarray*}
and $X,Y,Z$ stands for the cyclic permutations of $\text A,\text B,\text C$. In accordance with references \onlinecite{Wie03} and \onlinecite{Vid05} we define the pre-factors $E_\text A$, $E_\text B$, and $E_\text C$ in eq.~(\ref{free_energy}) as \emph{charging energies} of the individual QDs and $E_\text{AB}$, $E_\text{BC}$, and $E_\text{AC}$ as the \emph{electrostatic interdot coupling energies} between two QDs.

The charging energies, electrostatic coupling energies, and capacitances in eq.~(\ref{free_energy}) can be obtained from measurements, i.\,e.\ charge stability diagrams, and the conductance of the TQD in the non-linear regime. Equation~(\ref{free_energy}) only takes voltages explicitly into account, that are applied to the plunger gates ($V_\alpha$, $V_\beta$, and $V_\gamma$). It does not consider the detailed geometry of the TQD-device. Therefore, the free energy of a given configuration of the TQD is not completely determined by eq.~(\ref{free_energy}). Selection of a suitable zero point of the charge distribution scales a modeled stability diagram to fit measured data. This zero point might be defined as the charge on each QD at grounded plunger gates. The described model is suited rather for a qualitative than a quantitative analysis.

\begin{figure}[th]
\begin{center}
\epsfig{file=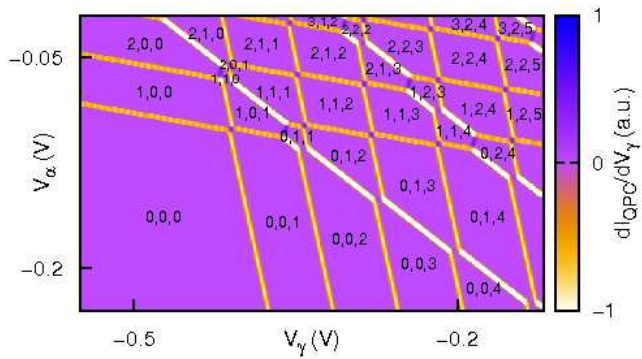, width=8.5cm}
\end{center} 
\vspace*{-5mm}
\caption{\label{fig3}(Color online) Numerically calculated ground state stability diagram of the TQD device shown in Fig.~\ref{fig1} for charging energies, electrostatic interdot coupling energies, and capacitances between QDs and plunger gates, similar to experimentally derived values. The color scale of the lines is chosen to simulate a possible measurement of the transconductance of the left QPC in Fig.~\ref{fig1} as a function of the plunger gate voltages $V_\alpha$ and $V_\gamma$, where $V_\gamma$ is modulated. The background color denotes zero transconductance. Stable charge configurations are labeled by triples of numbers ($N_\text A,N_\text B,N_\text C$).}
\end{figure}

Figure~\ref{fig3} 
shows a model stability diagram of a serial TQD calculated with eq.~(\ref{free_energy}) as explained above. Here, the y and x axes correspond to the plunger gate voltages $V_\alpha$ and $V_\gamma$ of the two outer QDs (compare Fig.~\ref{fig1}). All other gate voltages are kept constant. Lines mark borders of stable charge configurations ($N_\text A,N_\text B,N_\text C$).

The variable brightness of the lines in Fig.~\ref{fig3} simulates an experimental situation, where the left QPC in Fig.~\ref{fig1} would be used as charge detector. The brightness reflects the electrostatic coupling strength between the QDs and the left QPC. In a corresponding experiment the change of the current through the QPC in response to an amplitude modulation of the plunger gate voltage $V_\gamma$ will be measured. Adding the charge of one electron to the TQD while increasing $V_\gamma$ decreases the current through the detector QPC. This results in a \emph{charging line} of negative transconductance $\text{d}I_\text{QPC}/\text{d}V_\gamma$.

The spacing between charging lines belonging to a QD is approximately proportional to the charging energy $E_X$ of that QD.\cite{Wie03} The slope of a charging line is always negative and determined by the ratio of the respective couplings between a QD and the two plunger gates $\alpha$ and $\gamma$. Accordingly, the stability diagram of a TQD contains charging lines with three different main slopes. The nearly vertical charging lines indicate charging events of QD C, which couples strongly to plunger gate $\gamma$ but weakly to $\alpha$ (compare the gate layout in Fig.~\ref{fig1}). Lines corresponding to charging events of QD B (in the center) have a slope of $\text{d}V_\alpha/\text{d}V_\gamma=-1$, since we assumed equal capacitances between QD B and the two plunger gates $\alpha$ and $\gamma$. The predetermined symmetry properties result in charging lines with reciprocal slopes for QD A compared to C.

All crossings of two charging lines are avoided, because of the electrostatic interdot couplings. The result are pairs of two \emph{triple points}, each with three degenerate charge configurations. If well separated from charging lines of the third QD, the distance between the two triple points of a pair is proportional to the corresponding electrostatic interdot coupling energy $E_{XY}$. Since our model neglects quantum mechanical tunnel couplings these avoided crossings are of purely classical nature. The triple points of a pair are connected via a \emph{charge reconfiguration line}.\cite{Pet04} Along these lines with positive slopes an increase of $V_\gamma$ always causes a charge transfer between the three QDs with the center of charge moving away from the detector QPC, hence, resulting in a positive transconductance. The total charge of the TQD stays constant at the charge reconfiguration lines. 

Between lines of extremal transconductance the ground state charge configuration is stable and, hence, the transconductance is zero. The electrostatic interdot couplings can lead to the zig-zag course of charging and charge reconfiguration lines as clearly observable in the range of approximately six electron charges on the TQD. More complex behavior, including QCA-processes, is expected, where charging lines of all three slopes are close by, as will be discussed in section V.

The model stability diagram in Fig.~\ref{fig3} shows the situation expected for a TQD in the few electron regime. The lack of charging lines in the lower left corner of the figure indicates that here, the TQD is uncharged. Along the horizontal (vertical) axis QD C (A) is charged electron by electron. The plunger gate voltage $V_\beta$ is chosen such, that QD B can only be charged, if another QD is already occupied by at least one electron. However, increasing $V_\beta$ would shift the charging lines with slope $\text{d}V_\alpha/\text{d}V_\gamma=-1$ of QD B towards the lower left corner of the figure. The QDs A and C are separated by QD B and hence, have a relatively small mutual interdot coupling. This results in pairs of triple points being close to each other and charging lines that almost intersect (compare e.\,g.\ the transition between configurations $(0,0,1)\,\leftrightarrow\,(1,0,0)$ in Fig.~\ref{fig3}). In comparison, the electrostatic interdot couplings between neighboring QDs is much stronger resulting in a larger distance between triple points (e.\,g.\ see charge reconfiguration line between $(0,2,4)\,\leftrightarrow\,(1,1,4)$ in Fig.~\ref{fig3}). Note, that for the discussed model calculation we chose the coupling between QDs B and C to be smaller than that between QDs A and B.

Resonant tunneling transport of electrons through the TQD is only possible at \emph{quadruple points}, where four charge configurations are degenerate. However, as quadruple points are distinct points in a three-dimensional space, two-dimensional stability diagrams of a TQD containing quadruple points are rare. Since two charging lines can meet (but never cross) in one point of a stability diagram, a quadruple point of a TQD always represents a meeting point of two charging lines and two charge reconfiguration lines. A charge stability diagram in the direct vicinity of quadruple points contains up to eight triple points at four avoided crossings. A detailed discussion of this complex situation and comparisons with measured stability diagrams follows in section V.

In the case of a high degree of symmetry, i.\,e.\ equal interdot couplings $E_\text{AB}=E_\text{BC}=E_\text{CA}$, very rare hextuple points with six degenerate charge configurations are theoretically possible. Hextuple points involve the meeting of two charging lines and the crossing of two reconfiguration lines in one point. However, in our serial TQD geometry, where two electrostatic interdot coupling energies are larger than the third one, we would not expect to see such hextuple points.

Two important limits restrict the validity of the electrostatic model. The geometry of the electronic probability distribution inside a realistic TQD lacks perfect symmetry. It rather is a complicated function of applied gate voltages and the local disorder potential. This causes a non-linear gate voltage dependence of the capacitance matrix elements. It can result in a change of the distance between parallel charging lines or even in a continuous change of slopes of charging lines. In addition, our model neglects corrections caused by quantum mechanical tunneling. The classical avoided crossings are accompanied by quantum mechanical anticrossings, causing additional curvatures for interdot tunnel splittings comparable to the electrostatic interdot coupling energies.\cite{Hue05}

\section{Charge and transport measurements}

In this section we discuss the measured ground state stability diagram of the TQD structure in Fig.~\ref{fig1}. In the data presented we always use the left QPC (see Fig.~\ref{fig1}) as charge detector. \footnote{For the calculated stability diagrams shown in this article the color scales simulate a situation similar to presented data, where the left QPC is used as charge detector.} The differential conductance of the TQD is investigated within a range of the stability diagram that allows co-tunneling at strong enough tunnel couplings.

\begin{figure}[th]
\begin{center}
\epsfig{file=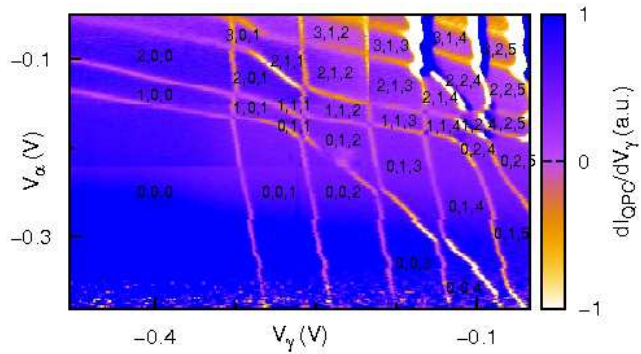, width=8.5cm}
\end{center} 
\caption{\label{fig4}(Color online) Measured charge stability diagram of the TQD device shown in Fig.~\ref{fig1} as a function of gate voltages $V_\alpha$ and $V_\gamma$. The color scale measures the transconductance of the left QPC in Fig.~\ref{fig1} as a function of the (modulated) plunger gate voltage $V_\gamma$. Stable charge configurations are denoted by triples of numbers ($N_\text A,N_\text B,N_\text C$). The graph is composed from several consecutive measurements explaining e.\,g.\ the horizontal line at $V_\alpha\sim-0.22\un{V}$.}
\end{figure}

The stability diagram in Fig.~\ref{fig4}
displays the transconductance $\text{d}I_\text{QPC}/\text{d}V_\gamma$ of the left QPC as a function of gate voltages $V_\alpha$ and $V_\gamma$ with constant voltages applied to all other gates.  Fig.~\ref{fig4} clearly demonstrates that a single nearby QPC is sensitive enough to monitor charging events in all three QDs. To measure transconductance using a lock-in amplifier, $V_\gamma$ is modulated with an amplitude of $\Delta V_\gamma\simeq 0.7\un{mV}$ at a frequency of $f=33\un{Hz}$. In all transconductance measurements shown in this article $V_\text{I}=-300\un{\mu V}$ is applied to ohmic contact I (compare Fig.~\ref{fig1}) in order to bias the QPC, if not stated otherwise. Measurements with a smaller bias voltage applied to the QPC assure that the linear response condition is fulfilled at $V_\text{I}=-300\un{\mu V}$. All other ohmic contacts besides contact I are grounded. 
The data in Fig.~\ref{fig4} features lines with three different main slopes as expected for a TQD (compare with Fig.~\ref{fig3}). Almost horizontal lines of minimum transconductance are charging lines of QD A. Likewise, almost vertical charging lines belong to QD C. The slope of $\text{d}V_\alpha/\text{d}V_\gamma\sim-1$ of the third kind of charging lines belongs to QD B and shows, that the electrostatic coupling strengths between QD B and the two plunger gates $\alpha$ and $\gamma$ have similar values. The absence of all three kinds of charging lines in the lower left corner of the stability diagram suggests that the TQD is completely uncharged in this area of Fig.~\ref{fig4}. 

However, the tunnel barriers of the QDs are larger for smaller gate voltages. In principle, charging lines can be invisible at very high tunnel barriers, if the charging process of a QD is slow compared to the time scales limiting the experiment. We ruled out the slow tunneling rate scenario by conducting careful test measurements, including different voltages applied to other gates than $\alpha$ and $\gamma$. We conclude therefore, that our TQD is really uncharged in the lower left corner of Fig.~\ref{fig4}.

\begin{figure}[th]
\begin{center}
\epsfig{file=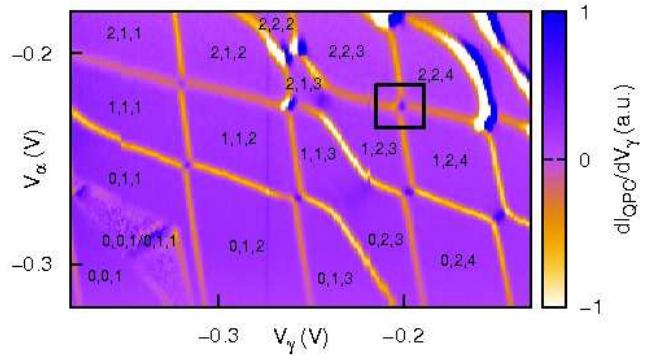, width=8.5cm}
\end{center} 
\caption{\label{fig5}(Color online) Expansion of a region of the TQD charge stability diagram in Fig.~\ref{fig4} for similar surface gate voltages. A black rectangle marks an area also marked by rectangles in Fig.~\ref{fig7}.}
\end{figure}

Figure~\ref{fig5}
expands a region of the stability diagram in Fig.~\ref{fig4} for similar gate voltages applied. The charge reconfiguration lines (of positive transconductance and positive slope) are well resolved. The electrostatic interdot coupling (proportional to the length of charge reconfiguration lines) between the distant QDs A and C is small compared to those between neighboring QDs. In addition, the interdot coupling between QDs A and B is larger than that between QDs C and B. This is reflected in the length of the charge reconfiguration lines between configurations $(0,1,3)\,\leftrightarrow\,(1,1,2)$ compared to $(0,2,3)\,\leftrightarrow\,(1,1,3)$. For few electrons charging the TQD, we find electrostatic interdot coupling energies of $E_\text{AB}\simeq680\un{\mu eV}$, $E_\text{BC}\simeq150\un{\mu eV}$ and $E_\text{AC}\simeq70\un{\mu eV}$. From the distances between charging lines we find charging energies of the order $E_\text{A}\simeq1.1\un{meV}$, $E_\text{B}\simeq2.0\un{meV}$ and $E_\text{C}\simeq1.0\un{meV}$. The conversion of gate voltages to energies is done with the help of non-linear transport measurements as will be discussed at the end of this section (compare Fig.~\ref{fig7}).

The model stability diagram in Fig.~\ref{fig3} was calculated for the energies and capacitances derived from the measured stability diagrams. We find a good agreement of the main features, including the mean distances between charging lines and triple points and the average slopes. A complete quantitative agreement is not expected, because of the limits of the model as discussed in section III. Moreover, the measured data reveal a spectrum of phenomena, not accounted for in the simple electrostatic model assuming a constant capacitance matrix. Some of these features are discussed below.

The interdot coupling between QDs A and C increases as $V_\gamma$ is increased (compare the charge reconfiguration lines in Fig.~\ref{fig5} between configurations $(0,1,3)\,\leftrightarrow\,(1,1,2)$ with $(0,2,4)\,\leftrightarrow\,(1,2,3)$ and $(0,2,5)\,\leftrightarrow\,(1,2,4)$). This effect, that can also be observed when increasing $V_\alpha$, can be explained by considering two aspects. First, the quantum mechanical tunnel coupling between neighboring QDs increases with increasing gate voltages, adding to the classical avoided crossing and causing a curvature of charging lines at the triple points. Second, the charge distribution and the position of the center of charges within all three QDs depend on the charge configuration.

The distance between the almost horizontal charging lines of QD A varies strongly. A detailed analysis yields a charging energy of QD A that is larger for the third electron than for the first or fourth electron (compare Fig.~\ref{fig4}). Such a strong effect implies a quite asymmetric confinement potential of QD A for the gate voltages applied. For instance, a situation where the first two electrons fit next to each other into an elongated QD, could explain the observation.

Charging lines belonging to different QDs have different brightness, reflecting the amplitude of the transconductance extrema. The brightness of a charging line is a linear function of its slope, diminishes at a larger width and is proportional to the electrostatic coupling between the QD and the detector QPC. The slope of a charging line determines the component of the line width parallel to the gate voltage $V_\gamma$, which is modulated and relevant for the transconductance measurement. The quantum life time of an electron in a QD directly influences the widths of the charging lines. Generally, charging lines widen as gate voltages are increased and the involved tunnel couplings grow. The charging line of QD B is brightest, because QD B is well decoupled from the leads and exhibits the longest quantum life time. The widths and brightness of a charge reconfiguration line depends on the corresponding interdot coupling rather than a quantum life time.

In the upper right corners of Figs.~\ref{fig4} and~\ref{fig5} charging lines turn into two parallel double lines, one with a large negative and one with a large positive transconductance. They are caused by current flowing through the TQD and the grounded contact III, branching off the current flowing from the biased contact I to the grounded contact II (compare Fig.~\ref{fig1}). Note, that the ohmic contacts have resistances in the order of a $R\sim500\,\Omega$, resulting in a small potential drop across the TQD. A current maximum at contact III causes a dip of the measured current at contact II. Hence, the transconductance $\text dI_\text{II}/\text{d}V_\gamma$ splits in a negative and a positive contribution, as observed. Comparison with Fig.~\ref{fig7} confirms the areas of enhanced transport. A finite current through the TQD along charging lines is caused by higher order tunneling processes. A detailed discussion of transport through the TQD follows at the end of this section.

Within the triangular area marked with $(0,0,1/0,1,1)$ in Fig.~\ref{fig5}, the TQD fluctuates between the two charge configurations. We expect this bistability to be generic for serial systems of more than two QDs in the limit of small tunnel rates as will be discussed in a separate publication.

Along the dark vertical line visible in Fig.~\ref{fig5} at $V_\gamma\simeq -0.272\un{V}$, the extrema of the transconductance appear to be more pronounced. This is caused by an internal switching in one of the measurement instruments and is not related to the TQD.

\begin{figure}[th]
\begin{center}
\epsfig{file=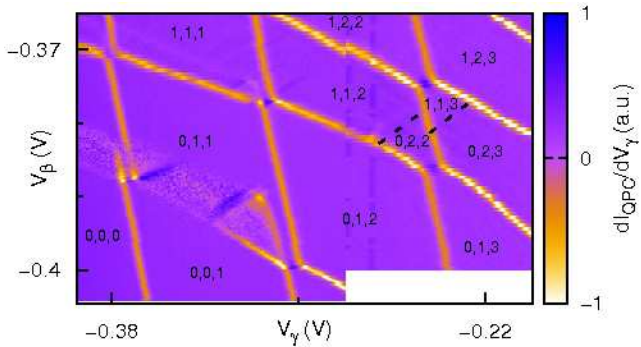, width=8.5cm}
\end{center}
\caption{\label{fig6}(Color online) TQD charge stability diagram as in Fig.~\ref{fig5}, but as a function of the plunger gate voltages $V_\beta$ and $V_\gamma$. The graph is composed out of three independent measurements. The white region was not investigated. Slight changes of the internal potential between the measurements lead to slightly imperfect seams between parts of the graph (e.\,g.\ at $V_\beta\simeq-0.37\,\text V; V_\gamma\simeq-0.28\,\text V$). Two dashed lines are a guide to the eye and follow very faint charge reconfiguration lines of positive transconductance. The bistable region observed in Fig.~\ref{fig5} is also visible.}
\end{figure}

\begin{figure}[th]
\centering
\subfigure{\label{transport_bias_0}\includegraphics{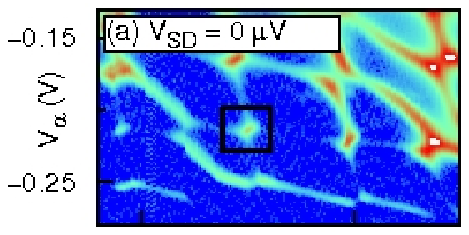}}%
\goodgap
\subfigure{\label{transport_bias_500}\includegraphics{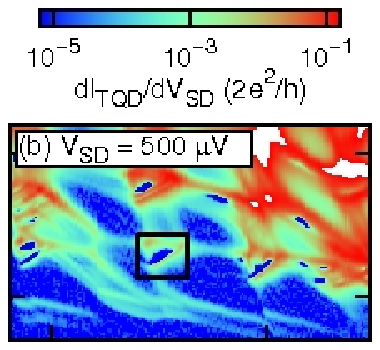}}\\%
\subfigure{\label{transport_bias_-150}\includegraphics{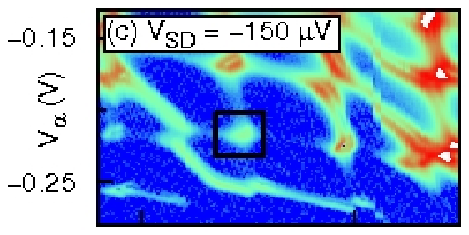}}%
\goodgap
\subfigure{\label{transport_bias_150}\includegraphics{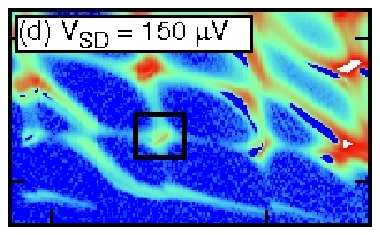}}\\%
\subfigure{\label{transport_bias_-300}\includegraphics{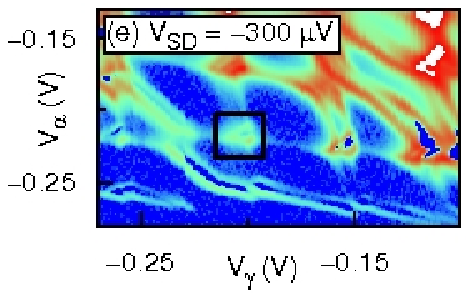}}%
\goodgap
\subfigure{\label{transport_bias_300}\includegraphics{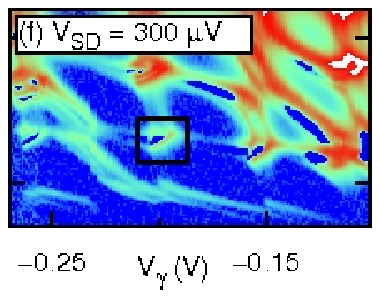}}\\%
\caption{\label{fig7}(Color online) The differential conductance $\text{d}I_\text{TQD}/\text{d}\VSD$ measured through the TQD plotted with a logarithmic color scale as a function of the plunger gate voltages $V_\alpha$ and $V_\gamma$. Voltages applied to other gates are as for Fig.~\ref{fig5}. Black rectangles mark the same region as in Fig.~\ref{fig5}. White color indicates a differential conductance exceeding the full range of the amplifier, small areas of the darkest blue color denote negative differential conductance. Applied bias voltages \VSD\ are indicated.}
\end{figure}

Fig.~\ref{fig6}
demonstrates the three dimensional nature of the TQD charge stability diagram. It shows a two-dimensional slice spanned by gate voltages $V_\beta$ and $V_\gamma$ perpendicular to that in Fig.~\ref{fig5}, which is spanned by $V_\alpha$ and $V_\gamma$. The two stability diagrams in Figs.~\ref{fig5} and \ref{fig6} together allow to determine all capacitances necessary for model calculations with eq.~\ref{free_energy}.
As before, charging lines with three different slopes can be identified, but two of the charging lines have similar slopes. They indicate a comparable ratio of couplings to both plunger gates $\beta$ and $\gamma$ for the two QDs A and B.
In addition, as will be discussed in section V, the data in Fig.~\ref{fig6} contain features characteristic for the regime in which all three QDs are nearly energetically degenerate.

Figure~\ref{fig7}
shows the differential conductance $\text{d}I_\text{TQD}/\text d\VSD$ of the TQD as a function of the plunger gate voltages $V_\alpha$ and $V_\gamma$ for various bias voltages $-0.3\un{mV}\le\VSD\le0.5\un{mV}$ between the source and drain contacts (II and III in Fig.~\ref{fig1}) of the TQD. The differential conductance is measured by means of lock-in technique with an ac modulation of $\Delta\VSD=20\un{\mu V}$ at a frequency of $f=33\un{Hz}$. For better comparison all gate voltages are identical for the transport measurements shown in Fig.~\ref{fig7} and the charge detection measurement displayed in Fig.~\ref{fig5}. Note that, compared to Fig.~\ref{fig5}, the area spanned by $V_\alpha$ and $V_\gamma$ is smaller for the transport measurements in Fig.~\ref{fig7}. It corresponds to the upper right corner of the stability diagram in Fig.~\ref{fig5}, where the tunnel couplings of the TQD are largest. The logarithmic color scale for the differential conductance in Fig.~\ref{fig7} overemphasizes very small currents through the TQD.

The differential conductance in linear response for $\VSD\simeq0$ (Fig.~\ref{fig7}a) is entirely caused by higher order tunneling processes as this area of the stability diagram does not contain quadruple points (compare discussion in section V). Alternatively, such higher order processes could be explained in a picture using molecular eigenstates\cite{Hue05} of a tri-atom. Here, we restrict ourself to the picture of higher order tunneling processes in the basis of single-dot eigenstates.

As the plunger gate voltages in Fig.~\ref{fig7}a are increased the tunnel barriers decrease and hence, the differential conductance increases. An exeption to this rule can be seen along the charging lines of slope $\text dV_\alpha/\text{d}V_\gamma\sim-1$ to the left of the black square in Fig.~\ref{fig7}a , along which the central QD B is in resonance with the 2DES in the leads. Here, the differential conductance is larger than along charging lines in the direct vicinity of the black square (for larger $V_\gamma$). Near the black square only QD A or C can be resonant with the chemical potentials of the leads. Here, if away from triple points, current through the TQD is caused by third order tunneling processes through two non-resonant QDs in series. In contrast, along the charging lines of QD B an electron can occupy a resonant state in QD B between two sequential co-tunneling processes. Transport via such two successive second order processes is highly enhanced compared to transport involving one resonant first order and one third order tunneling process.\cite{Fra00}
 
For $\VSD\ne0$ the charging lines in Fig.~\ref{fig7} split into two parallel lines corresponding to two different resonances of one of the QDs with the source or drain potential in the two leads. For the same reason triple points turn into triangles. Similar as for a DQD,\cite{Wie03} the size of such a triangle or the distance between parallel double lines can be used to determine the conversion factors between gate voltages and the energy scales. These conversion factors are needed in order to calibrate the ground state stability diagrams and determine the charging energies and electrostatic interdot coupling energies of the QDs.

\section{Vicinity of quadruple points}

\begin{figure}[tb!]
\begin{center}
\epsfig{file=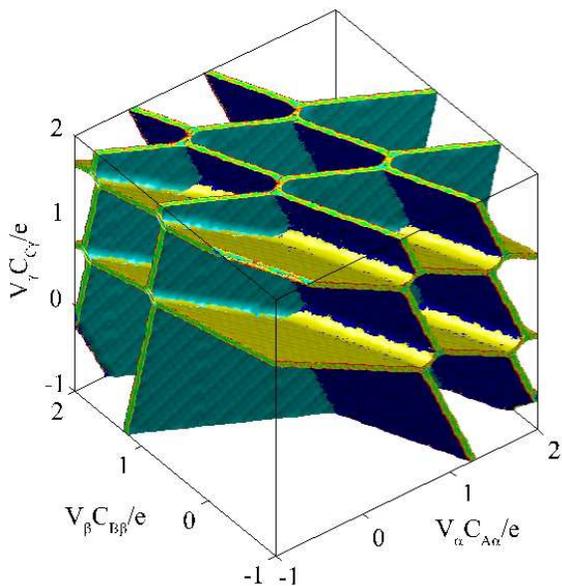, width=7.5cm}
\end{center}
\caption{\label{fig8}(Color online) Three-dimensional charging diagram, or beehive diagram, showing stable configurations of the triple-dot system as a function of the three plunger-gate biases, normalised by their capacitive couplings to the \emph{closest} dot.  The front and bottom region of the structure corresponds to the configuration $(0,0,0)$. On each of three visible end-faces resembles a two-dimensional honeycomb-like diagram, although more complicated diagrams can be seen at other slices, as shown in Fig.~\ref{fig9}.}
\end{figure}

Because of the presence of three discrete charges in the triple-dot system, the charge-configuration diagrams are formally defined in a three-dimensional phase space. By analogy with the two-dimensional honeycomb diagrams seen for double-dot structures, we term this three-dimensional charge configuration diagram a beehive diagram, and a calculated example is shown in Fig.~\ref{fig8}.
In this case, we have deliberately chosen a capacitance matrix regime with minimal cross-coupling, so that the planes of the three visible end-faces show honeycomb-like charging diagrams.

In the remainder of this article we focus on an area of the stability diagram where all three QDs of the TQD are close to being resonant with the chemical potential in the adjacent 2DES, that is where charging lines of all three QDs are close by.

In the case of a DQD avoided crossings of any two charging lines always result in two triple points enclosed by four different areas of stable charge configurations, since two QDs are charged each by up to one additional electron ($2^2=4$). In a TQD charging lines of three different slopes (belonging to the three QDs) exist. If two of them meet in a two-dimensional stability diagram they form triple points just as it is the case for a DQD (see Fig.~\ref{fig3}). In a three-dimensional stability diagram of a TQD, e.\,g.\ spanned by the plunger gates $\alpha$, $\beta$, and $\gamma$, charging lines turn into planes and triple points turn into lines, (compare Fig.~\ref{fig8}). In a region where charging planes of all three QDs meet, each QD can be charged by one additional electron. This results in $2^3=8$ possible charge configurations, surrounding four avoided crossings with eight triple lines. Such a three dimensional structure contains four quadruple points, where two charging planes and two charge reconfiguration planes meet. Only at these quadruple points is transport by sequential resonant tunneling of electrons through a serial TQD possible.

\begin{figure*}[th]
\setlength{\unitlength}{1bp}%
\begin{center}
\hspace{-15mm}
\resizebox{0.87\textwidth}{!}{
\begin{picture}(400,250)
\put(360,162) {
\rotatebox{90}{\rm model calculations}
}
\put(39,200){
    \includegraphics[]{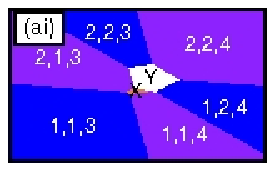}%
}
\put(119,200) {
    \includegraphics[]{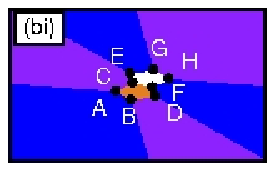}%
}
\put(199,200) {
    \includegraphics[]{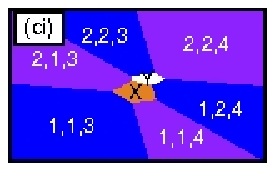}%
}
\put(279,200) {
    \includegraphics[]{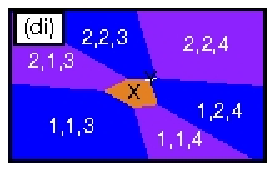}%
}
\put(39,153) {
    \includegraphics[]{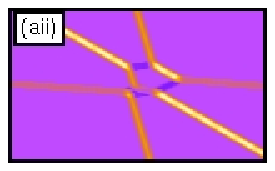}%
}
\put(119,153) {
    \includegraphics[]{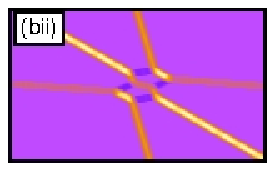}%
}
\put(199,153) {
    \includegraphics[]{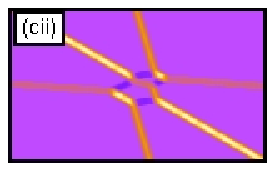}%
}
\put(279,153) {
    \includegraphics[]{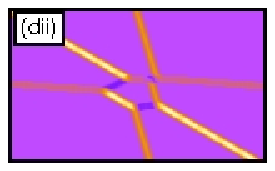}%
}
\put(39,106) {
    \includegraphics[]{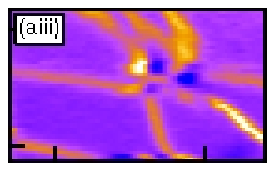}
}
\put(119,106) {
    \includegraphics[]{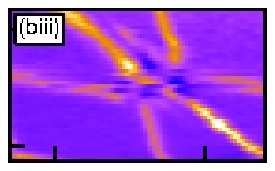}%
}
\put(199,106) {
    \includegraphics[]{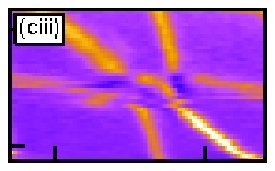}%
}
\put(279,106) {
    \includegraphics[]{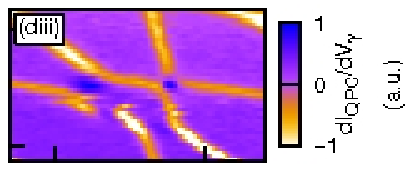}%
}
\put(0,59) {
    \includegraphics[]{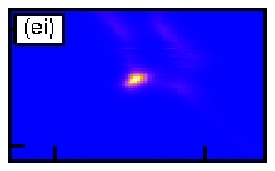}
}
\put(80,59) {
    \includegraphics[]{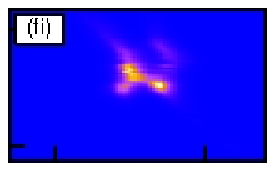}
}
\put(160,59) {
    \includegraphics[]{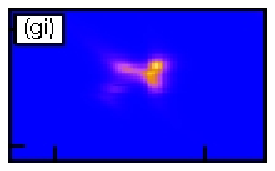}
}
\put(240,59) {
    \includegraphics[]{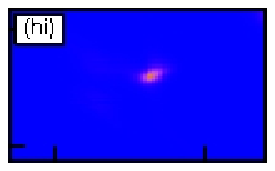}
}
\put(320,59) {
    \includegraphics[]{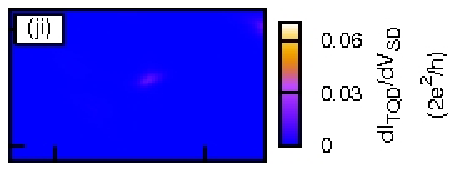}
}
\put(0,12) {
    \includegraphics[]{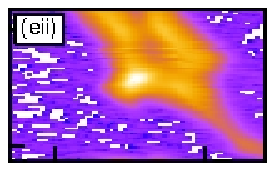}%
}
\put(80,12) {
    \includegraphics[]{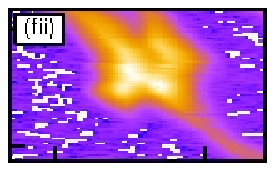}%
}
\put(160,12) {
    \includegraphics[]{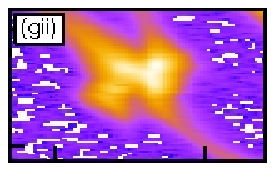}%
}
\put(240,12) {
    \includegraphics[]{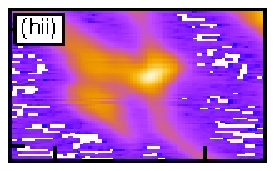}%
}
\put(320,12) {
    \includegraphics[]{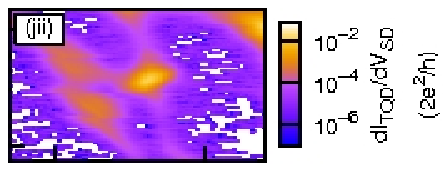}%
}
\put(7,3) {
    \includegraphics[scale=0.35]{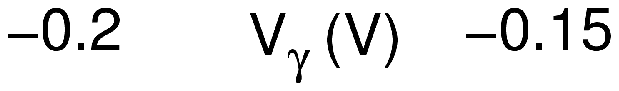}%
}
\put(-16,15) {
    \includegraphics[scale=0.35]{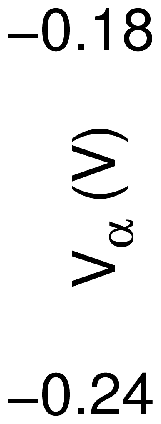}%
}
\end{picture}
}
\end{center}
\caption{\label{fig9}(Color online) Expansion of a region of the stability diagram of the TQD as a function of the plunger gate voltages $V_\alpha$ and $V_\gamma$. Here, charging lines of all three QDs are close by. $V_\beta$ is increased in steps of $2\un{mV}$ from (a) to (d) or from (e) to (i). The upper two rows show identical results from model calculations in two different representations. Triples of numbers denote stable charge configurations, where $\text X=1,2,3$ and $\text Y=2,1,4$. The third row shows transconductance measurements as e.\,g.\ in Fig.~\ref{fig5}. The two lowest rows display identical differential conductance measurements of the TQD with a linear (upper) and a logarithmic (lowest row) color scale (see main text for more explanations). The voltage ranges of $V_\alpha$ and $V_\gamma$ are identical for all subplots and shown in the left corner.}
\end{figure*}

To more clearly examine the three-dimensional stability diagram in an experimentally accesible fashion, we study a series of parallel two-dimensional slices through the beehive diagram. We concentrate on regions of high degeneracy and use the terms appropriate for two dimensions as triple points and charging lines. In Fig.~\ref{fig9}
such measurements are plotted as a function of $V_\alpha$ and $V_\gamma$ and compared with model calculations. The voltage ranges of $V_\alpha$ and $V_\gamma$ are identical for all subplots. The third plunger gate voltage $V_\beta$ is increased in steps of $2\un{mV}$ between $-396\un{mV}\le V_\beta\le-390\un{mV}$ from (a) to (d). The two bottom rows of Fig.~\ref{fig9} show conductance measurements, that will be discussed later. The middle row plots the transconductance of QPC charge detection measurements of comparable regions of the stability diagram. The two upper rows feature identical numerical calculations according to our model, displayed with two different methods.

The transconductance is measured with the left QPC with a bias voltage of $V_\text{II}=-300\un{\mu V}$ applied to contact II (compare Fig.~\ref{fig1}). This bias voltage also causes current through the TQD at certain places of the stability diagram. This is proven by the conductance measurements in linear response plotted in the two bottom rows of Fig.~\ref{fig9}. Thus, the finite bias applied to contact II generates additional features in the transconductance measurements. These include extra lines, e.\,g.\ a line with slope $-1$ within the $(2,2,3)$ region in Fig.~\ref{fig9}\,(biii), and gaps that interrupt lines, e.\,g.\ on the bottom left of Fig.~\ref{fig9}\,(diii). A small shift between the position of features in the transconductance data as compared to the conductance data can partly be explained by the applied biases, but could as well be caused by potential drifts during the time gap between these experiments. For a rough compensation the conductance measurements in the lowest two rows of Fig.~\ref{fig9} are horizontally shifted by $\Delta V_\beta=1\un{mV}$. Unfortunately, no charge detection measurements with the TQD left unbiased exist so far. Nevertheless, all features of the model calculations (first two rows of Fig.~\ref{fig9}) are clearly seen in the measured transconductance data (third row of Fig.~\ref{fig9}).

Charge configurations, identified from a larger area stability diagram, are depicted in the first row of the model stability diagrams
of Fig.~\ref{fig9}, where configurations X and Y equal $(1,2,3)$ and $(2,1,4)$, respectively. Placing the stability diagrams from Fig.~\ref{fig9} (a) to (d) above each other, with distances corresponding to $V_\beta$, results in a three-dimensional section of the stability diagram. Regions X and Y are similar to irregular pentangular-based pyramids inverted with respect to each other. (as can be seen in Fig.~\ref{fig9}). The tips of the pyramids are oriented in approximately opposite direction from each other at two quadruple points. 

For the approximately symmetric case shown in Fig.~\ref{fig9}(b), the central charging line with slope $\text{d}V_\alpha/\text{d}V_\gamma\sim-1$ (belonging to QD B) shows a zig-zag behavior resulting in four triple points. In addition, the other two charging lines both contain a short segment parallel to the central charging line with $\text{d}V_\alpha/\text{d}V_\gamma\sim-1$. These features can be roughly explained as follows: Imagine the central charging line would be absent. Then we were left with one avoided crossing, where four lines end in two triple points. The central line, once added, repels the other four charging lines and four new avoided crossings occur, resulting in the observed geometry with eight triple points. 

In the following discussion we use a notation that subtracts the common charge state $(1,1,3)$ to be left with configurations of type $(u,v,w)$ with $u,v,w=0,1$. In the most symmetric case the eight triple points then involve the following degenerate charge configurations (compare labels in Fig.~\ref{fig9}\,(bi))
\begin{eqnarray*}
\text{TP}_\text{A}&:& (0,0,0) \leftrightarrow (1,0,0) \leftrightarrow (0,1,0)\\
\text{TP}_\text{B}&:& (0,1,0) \leftrightarrow (0,0,1) \leftrightarrow (0,0,0)\\
\text{TP}_\text{C}&:& (1,0,1) \leftrightarrow (1,0,0) \leftrightarrow (0,1,0)\\
\text{TP}_\text{D}&:& (0,1,1) \leftrightarrow (0,1,0) \leftrightarrow (0,0,1)\\
\text{TP}_\text{E}&:& (1,1,0) \leftrightarrow (1,0,1) \leftrightarrow (1,0,0)\\
\text{TP}_\text{F}&:& (1,0,1) \leftrightarrow (0,1,1) \leftrightarrow (0,1,0)\\
\text{TP}_\text{G}&:& (1,1,1) \leftrightarrow (1,1,0) \leftrightarrow (1,0,1)\\
\text{TP}_\text{H}&:& (1,0,1) \leftrightarrow (0,1,1) \leftrightarrow (1,1,1).
\end{eqnarray*}
At TP$_\text A$ either QD A or B can be resonantly occupied by an additional electron from the leads, but the occupation of QD C is energetically forbidden. Hence, at TP$_\text A$ sequential tunneling of an electron through the TQD requires one co-tunneling process via an energetically forbidden state in QD C. Similarly, transport at any other triple point requires one second order tunneling process. While TP$_\text A$ and TP$_\text B$ allow sequential co-tunneling of an electron, TP$_\text G$ and TP$_\text H$ allow sequential co-tunneling of a hole. Second order transport through the other four triple points involves two particles.

Where two charge reconfiguration lines (blue) with positive slopes and positive transconductance meet, two triple points combine to a quadruple point. As a function of $V_\beta$ this is possible for triple points $\text{TP}_\text{A}$ and $\text{TP}_\text{B}$ (close to the situation in Fig.~\ref{fig9}\,(a)), triple points $\text{TP}_\text{G}$ and $\text{TP}_\text{H}$ (between Fig.~\ref{fig9}\,(c) and (d)), triple points $\text{TP}_\text{C}$ and $\text{TP}_\text{E}$ (close to the situation in Fig.~\ref{fig9}\,(c)), and triple points $\text{TP}_\text{D}$ and $\text{TP}_\text{F}$ (between Fig.~\ref{fig9}\,(a) and (b)). The resulting quadruple points involve the following degenerate charge configurations
\begin{eqnarray*}
\text{QP}_\text{AB}&:& (0,0,0) \leftrightarrow (1,0,0) \leftrightarrow (0,1,0) \leftrightarrow (0,0,1)\\
\text{QP}_\text{CE}&:& (1,0,0) \leftrightarrow (0,1,0) \leftrightarrow (1,1,0) \leftrightarrow (1,0,1)\\
\text{QP}_\text{DF}&:& (0,0,1) \leftrightarrow (1,0,1) \leftrightarrow (0,1,1) \leftrightarrow (0,1,0)\\
\text{QP}_\text{GH}&:& (1,1,1) \leftrightarrow (1,1,0) \leftrightarrow (1,0,1) \leftrightarrow (0,1,1).
\end{eqnarray*}
At these four quadruple points resonant tunneling through the TQD is possible, e.\,g.\ at QP$_\text{AB}$ an electron can sequentially tunnel from the left lead into QD A, QD B, QD C, and then escape to the right lead (or vice versa). Quadruple points $\text{Q}_\text{AB}$ and $\text{Q}_\text{GH}$, respectively, allow sequential tunneling of an electron versus a hole through the TQD, similar to triple points in a DQD. However, the nature of transport at quadruple points $\text{Q}_\text{DF}$ and $\text{Q}_\text{CE}$ can not be described by one electron or hole tunneling through the TQD, but involves two particles (electrons or holes). This extends the possibilities in a DQD, where only electron- or hole-like transport is possible.

As a function of $V_\beta$ the pairs of triple points $\text{TP}_\text{C}$ and $\text{TP}_\text{E}$ as well as $\text{TP}_\text{D}$ and $\text{TP}_\text{F}$ meet in the corresponding quadruple points, respectively, and then diverge again. During this process at the quadruple point one resonant charge configuration is exchanged between a pair of triple points, resulting in the modified triple points
\begin{eqnarray*}
\text{TP}_\text{C}^\prime&:& (1,0,0) \leftrightarrow (0,1,0) \leftrightarrow (1,1,0)\\
\text{TP}_\text{E}^\prime&:& (0,1,0) \leftrightarrow (1,1,0) \leftrightarrow (1,0,1)
\end{eqnarray*}
\begin{eqnarray*}
\text{TP}_\text{D}^\prime&:& (0,1,0) \leftrightarrow (0,0,1) \leftrightarrow (1,0,1)\\
\text{TP}_\text{F}^\prime&:& (0,0,1) \leftrightarrow (1,0,1) \leftrightarrow (0,1,1).
\end{eqnarray*}
Compared to the approximately asymmetric case in Fig.~\ref{fig9}(b)
in Fig.~\ref{fig9}(a) $\text{TP}_\text{D}$ and $\text{TP}_\text{F}$ are replaced by $\text{TP}_\text{D}^\prime$ and $\text{TP}_\text{F}^\prime$ and in Fig.~\ref{fig9}(d) $\text{TP}_\text{C}$ and $\text{TP}_\text{E}$ are replaced by $\text{TP}_\text{C}^\prime$ and $\text{TP}_\text{E}^\prime$.

Pairs of quadruple points as $\text{QP}_\text{AB}\,\leftrightarrow\,\text{QP}_\text{GH}$ as well as $\text{QP}_\text{CE}\,\leftrightarrow\,\text{QP}_\text{DF}$ show electron-hole symmetry, respectively. The same is true for triple points, e.\,g.\ $\text{TP}_\text{A}\,\leftrightarrow\,\text{TP}_\text{H}$. In addition, after subtraction of the common charge state $(1,1,3)$, triple points are pairwise point symmetric in respect to the central QD B regarding their charge occupation, e.\,g.\ $\text{TP}_\text{A}\,\leftrightarrow\,\text{TP}_\text{B}$.

The regions X and Y of stable charge configurations read $\text X=(0,1,0)$ and $\text Y=(1,0,1)$ after subtraction of the common charge state $(1,1,3)$. Crossing the line of minimum transconductance, separating these two areas, from X towards Y involves adding a charge to QD A (or C). However, this is only possible via a QCA-process, where simultaneously one electron is pushed from the central QD B into QD C (or A). This is a combination of charging one QD and a charge reconfiguration between the other two QDs. Therefore, the slope of the QCA-line between regions X and Y is determined by the combination of the two processes involved and differs from all other charging line slopes in the stability diagram.

Let us now consider the reverse process, which involves crossing the QCA-line from Y towards X. During this second order tunneling process an electron leaves QD C (or A) and simultaneously pulls another electron from QD A (or C) into the  central QD B. Interestingly, the combination of both processes (crossing the QCA-line for- and backwards) can result into transport of one electron through the TQD via two successive second order tunneling processes, similar as along the charging line of QD B.

Second order tunneling processes that preserve charge are usually called co-tunneling processes. The QCA processes described above are not charge preserving, but second order. Hence, we refer to these as \emph{QCA-co-tunneling processes}.

Note, that an equivalent situation to that shown in Fig.~\ref{fig9} occurs in the upper right quarter of the stability diagram in Fig.~\ref{fig6}, but here, as a function of the plunger gate voltages $V_\beta$ and $V_\gamma$. In Fig.~\ref{fig6} two charge transfer lines are retraced by dashed lines as a guide to the eyes. Because QDs A and B feature comparable electrostatic couplings to both plunger gates $\beta$ and $\gamma$, some of the triple points are hardly seen in Fig.~\ref{fig6} (see also above discussion of Fig.~\ref{fig6}).

The lowest two rows of Fig.~\ref{fig9} display the conductance of the TQD plotted both with a linear (second lowest row) and with a logarithmic (bottom row) color scale. The conductance is measured in the linear response regime and for zero dc bias on all ohmic contacts. Comparison of the logarithmic conductance representation with the model calculations in Fig.~\ref{fig9} shows, that along the charging lines belonging to the central QD B and at the QCA-line a small current flows through the TQD. Both kind of charging lines are distinguished, because they allow transport through the TQD via two successive second order tunneling processes. As discussed above, along the other charging lines of QDs A and C only third order tunneling processes can cause transport. Accordingly, no current can be observed along charging lines belonging to QDs A and C.

The linear representation of the conductance through the TQD (second lowest row in Fig.~\ref{fig9}) reveals distinct current maxima at quadruple points, and triple points near quadruple points. A detailed comparison with the model calculations in Fig.~\ref{fig9} suggests, that the conductance maxima in Fig.~\ref{fig9}(ei) and~\ref{fig9}(ji) are very close to the quadruple points $\text{QP}_\text{AB}$ and $\text{QP}_\text{GH}$, respectively. Fig.~\ref{fig9}(fi) and~\ref{fig9}(gi) each show four bright maxima. The lower left one in Fig.~\ref{fig9}(fi) corresponds to the quadruple point also seen in Fig.~\ref{fig9}(ei). The upper left two maxima mark the triple points $\text{TP}_\text{C}$ and $\text{TP}_\text{E}$ in close vicinity of $\text{QP}_\text{CE}$, and the lower right maximum is close to $\text{QP}_\text{DF}$. Figure~\ref{fig9}(gi) can be described accordingly.

Strikingly, the QCA-line connecting $\text{TP}_\text{C}$ and $\text{TP}_\text{F}$ in Fig.~\ref{fig9}(bi) is also visible as a line of minimal transconductance in Fig.~\ref{fig9}(fi) and~\ref{fig9}(gi). On the other hand, the current flowing at the charging line of QD B is too small to be seen in the linear representation. Both transport channels involve two successive second order tunneling processes. Still, the QCA-line near quadruple points shows a larger conductance than the charging lines of the central QD B. This suggests that QCA-co-tunneling processes, with two particles simultaneously moving, result in a larger tunneling probability than regular second order co-tunneling processes, that can be explained in a one-particle picture. The origin of this phenomenon lies in the electrostatic interaction between two electrons tunneling simultaneously and will be discussed in a separate publication.

Finally, we would like to note that spin blockade of transport in linear response through a TQD is expected for certain quadruple point configurations. It was not observed in the regime treated in Fig.~\ref{fig9}. This can in part be explained by the special configuration, where up to three electrons are added to the charge state $(1,1,3)$, but at most one electron to each QD. In the configuration $(1,1,3)$, each QD already has a spin $1/2$. After an extra electron charge has been added to one of the QDs, this QD has zero spin. This implies that this QD can now provide an electron with arbitrary spin (up or down) to tunnel to an adjacent QD. Hence, a full spin blockade is not expected for the region of the stability diagram discussed in Fig.~\ref{fig9}.

\section{Conclusion}

In summary, we have realized a lithographically defined serial triple quantum dot that can be tuned to contain any number of electrons between zero and about ten in various configurations. Quantum point contacts as integrated charge sensors allow to determine the exact number of electrons charging each of the quantum dots. We have studied the ground-state stability diagram of the triple quantum dot in close vicinity to quadruple points where four different charge configurations are energetically degenerate. In this regime, quantum cellular automata processes are observed among other features, adding to the physics that can be found in double quantum dots. A simple electrostatic model, that can easily be scaled to larger structures containing more than three quantum dots, is compared with our data. A detailed discussion of the conductance near quadruple points reveals several kinds of tunneling processes. Quantum cellular automata co-tunneling processes lead to an enhanced conductance at only two-fold degeneracy of the triple quantum dot. The excellent control of charge configurations and tunnel couplings achieved in this triple quantum dot now opens the possibility to study coherent dynamics, e.\,g.\ of charge and possible spin transfer in such a complex quantum system.

\begin{acknowledgments}

We thank V.\ Khrappay, S.\ Kehrein, A.\ Sachrajda, and J.\,J.\ Greentree for fruitful discussions. The authors acknowledge financial support by the Deutsche Forschungsgemeinschaft via SFB 631 and by the German excellence initiative via the cluster ``Nanosystems Initiative Munich (NIM)", and the Bundesministerium f\"ur Bildung und Forschung via DIP-H.2.1. Additionally, ADG and LCLH are supported by the Australian Research Council, the Australian Government, and the U.\,S.\ National Security Agency (NSA), Advanced Research and Development Activity (ARDA), and the Army Research Office (ARO) under Contract No.\ W911NF-04-1-0290. LG acknowledges support from NSERC and the Canadian Institute for Advanced Research.
\end{acknowledgments}


\end{document}